\documentclass[
superscriptaddress,
twocolumn,
showpacs,preprintnumbers,
amsmath,amssymb,
aps,
]{revtex4-1}

\usepackage{graphicx}
\usepackage{dcolumn}
\usepackage{bm}
\usepackage[mathlines]{lineno}
\usepackage{hyperref}
\hypersetup{
    colorlinks=true,
    linkcolor=blue,
    filecolor=blue,
    urlcolor=blue,
    citecolor=blue,
}

\begin{document}

\title{Spin-Orbital States and Strong Antiferromagnetism of Layered  Eu$_2$SrFe$_2$O$_6$ and Sr$_3$Fe$_2$O$_4$Cl$_2$}

\author{Di Lu}
\affiliation{Laboratory for Computational Physical Sciences (MOE),
	State Key Laboratory of Surface Physics, and Department of Physics,
	Fudan University, Shanghai 200433, China}
\affiliation{Shanghai Qi Zhi Institute, Shanghai 200232, China}

\author{Ke Yang}
\affiliation{College of Science, University of Shanghai for Science and Technology,
       Shanghai 200093, China}
       \affiliation{Laboratory for Computational Physical Sciences (MOE),
	State Key Laboratory of Surface Physics, and Department of Physics,
	Fudan University, Shanghai 200433, China}

\author{Lu Liu}
\affiliation{Laboratory for Computational Physical Sciences (MOE),
	State Key Laboratory of Surface Physics, and Department of Physics,
	Fudan University, Shanghai 200433, China}
\affiliation{Shanghai Qi Zhi Institute, Shanghai 200232, China}

\author{Guangyu Wang}
\affiliation{Laboratory for Computational Physical Sciences (MOE),
	State Key Laboratory of Surface Physics, and Department of Physics,
	Fudan University, Shanghai 200433, China}
\affiliation{Shanghai Qi Zhi Institute, Shanghai 200232, China}

\author{Hua Wu}
\email{wuh@fudan.edu.cn}
\affiliation{Laboratory for Computational Physical Sciences (MOE),
	State Key Laboratory of Surface Physics, and Department of Physics,
	Fudan University, Shanghai 200433, China}
\affiliation{Shanghai Qi Zhi Institute, Shanghai 200232, China}
\affiliation{Collaborative Innovation Center of Advanced Microstructures,
	Nanjing 210093, China}

\begin{abstract}

The insulating iron compounds Eu$_2$SrFe$_2$O$_6$ and Sr$_3$Fe$_2$O$_4$Cl$_2$ have high-temperature antiferromagnetic (AF) order despite their different layered structures. Here we carry out density functional calculations and Monte Carlo simulations to study their electronic structures and magnetic properties aided with analyses of the crystal field, magnetic anisotropy, and superexchange. We find that both compounds are Mott insulators and in the high-spin (HS) Fe$^{2+}$ state ($S$ = 2) accompanied by the weakened crystal field. Although they have different local coordination and crystal fields, the Fe$^{2+}$ ions have the same level sequence and ground-state configuration $(3z^2-r^2)^2(xz,yz)^2(xy)^1(x^2-y^2)^1$. Then, the multiorbital superexchange produces strong AF couplings, and the $(3z^2-r^2)/(xz,yz)$ mixing via the spin-orbit coupling (SOC) yields a small in-plane orbital moment and anisotropy. Indeed, by tracing a set of different spin-orbital states, our density functional calculations confirm the strong AF couplings and the easy planar magnetization for both compounds. Moreover, using the derived magnetic parameters, our Monte Carlo simulations give the N\'eel temperature $T_{\rm N}$ = 420 K (372 K) for the former (the latter), which well reproduce the experimental results. Therefore, the present study provides a unified picture for Eu$_2$SrFe$_2$O$_6$ and Sr$_3$Fe$_2$O$_4$Cl$_2$ concerning their electronic and magnetic properties.

\end{abstract}

\maketitle

\section{Introduction}
Recently, layered $A$FeO$_2$ ($A$=Sr,Ba,Ca) oxides with a planar FeO$_2$ square lattice were synthesized from $A$FeO$_3$ perovskites via the topochemical reduction~\cite{tsujimoto2007,tassel2008,tassel2009,yamamoto2010}. They are quite interesting for being an above-room-temperature AF insulator despite their layered structures. Their electronic and magnetic properties are intimately correlated with the spin-orbital states of the constituent Fe ions. In addition, they could be further tuned by doping, pressure, and strain, $etc$. For example, SrFeO$_2$ was reported to undergo a pressure-induced spin crossover and an insulator-metal transition, even forming a ferromagnetic (FM) half-metallic state of lasting interest~\cite{kawakami2009}.

\begin{figure}
    \centering
    \includegraphics[width=8cm]{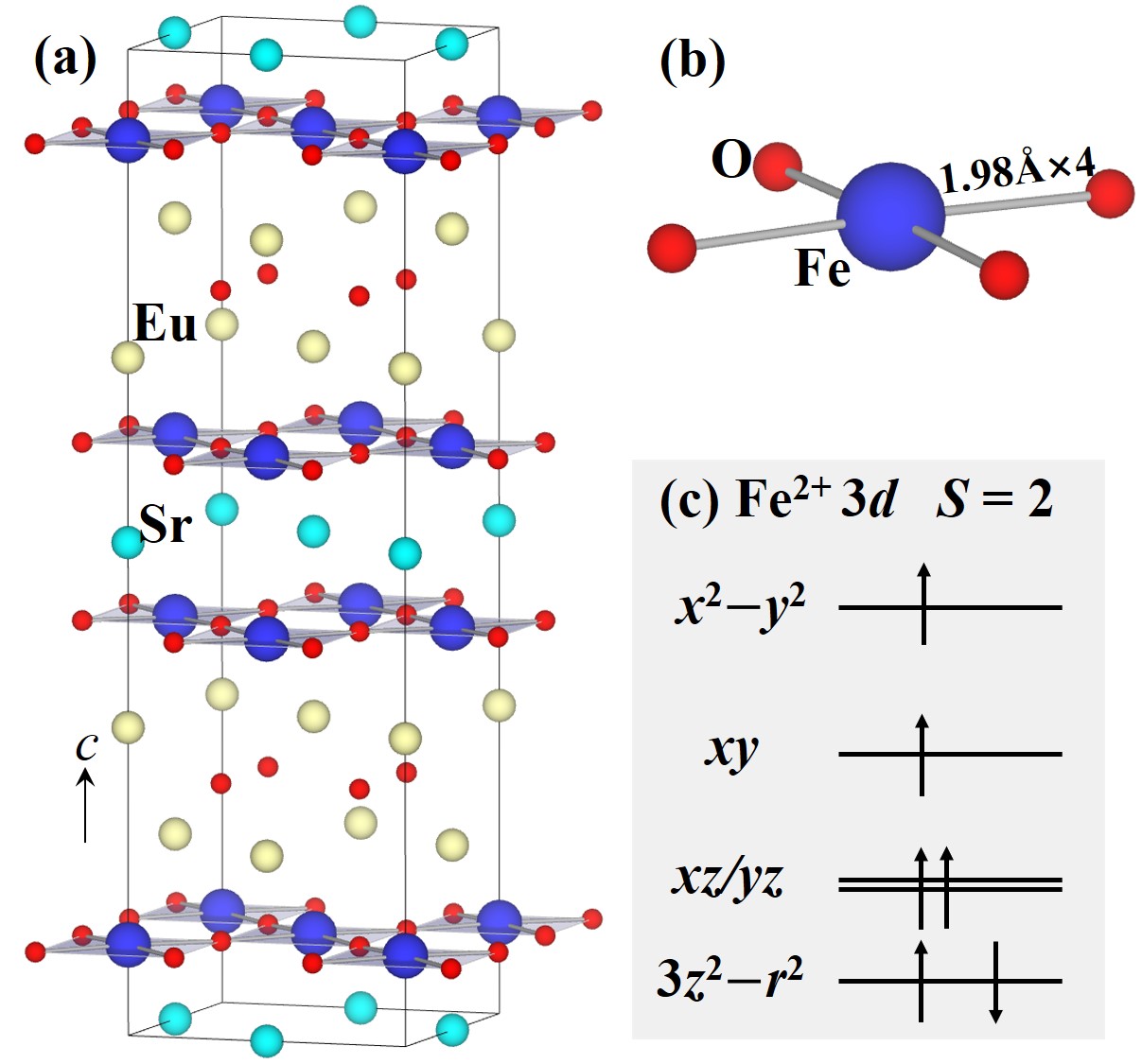}
    \caption{(a) $\sqrt{2}\times\sqrt{2}\times1$ structure of the bilayered Eu$_2$SrFe$_2$O$_6$ with (b) the local $4$-fold planar FeO$_4$ square. (c) Schematic crystal field level diagrams of the Fe$^{2+}$ $S$ = 2 state.}
    \label{Fig. 1}
\end{figure}

Low-dimensional magnets have long been of considerable interest. The bilayered Sr$_3$$B_2$O$_4$Cl$_2$ ($B$ = Fe,Co)~\cite{dixon2010,dixon2011,romero2012} compounds were also prepared recently. Recently, another bilayered Eu$_2$SrFe$_2$O$_6$ with $ab$ planar magnetism was synthesized from the $A$-site-ordered layered perovskite Eu$_2$SrFe$_2$O$_7$~\cite{lopez2020}. The temperature dependence of the Mossbauer spectrum and neutron powder diffraction data indicated the onset of G-AF (both intra- and interplane AF couplings) magnetic order at a N\'eel temperature $T_{\rm N}$ $\approx$ 390-404 K for Eu$_2$SrFe$_2$O$_6$ with a HS Fe$^{2+}$ configuration. However, with the same HS Fe$^{2+}$ configuration and a G-AF magnetic order below $T_{\rm N}$ = 378 K, Sr$_3$Fe$_2$O$_4$Cl$_2$ was reported to have a $c$-axis magnetization~\cite{dixon2010}. Naturally, several questions arise: why do these two similar bilayered materials with the same HS Fe$^{2+}$ charge-spin state have such a contrasting magnetic orientation? Are their orbital states much different to account for this discrepancy? Is there a unified picture to understand their electronic and magnetic properties?

\begin{figure}
    \centering
    \includegraphics[width=8cm]{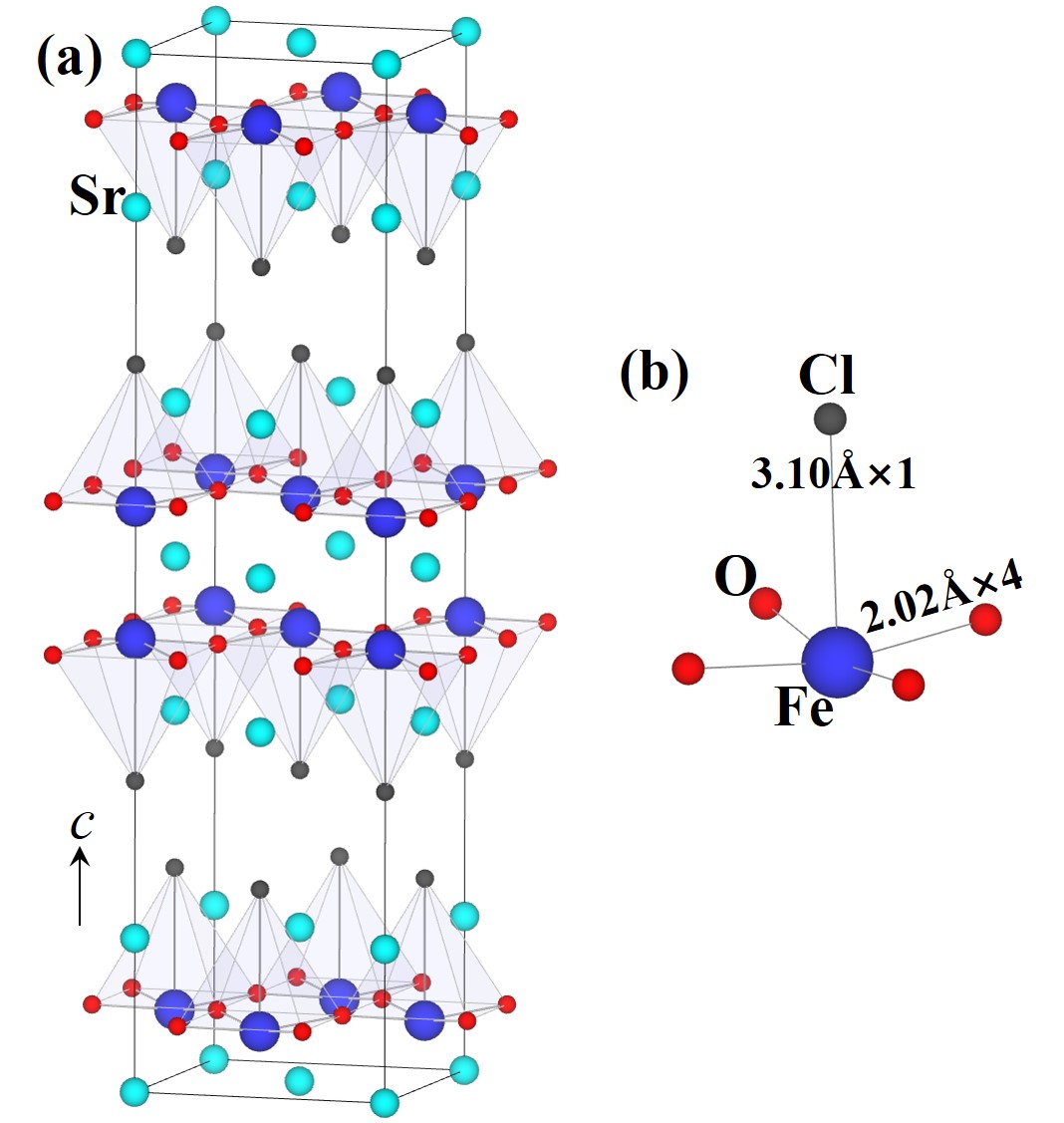}
    \caption{(a) $\sqrt{2}\times\sqrt{2}\times1$ structure of the bilayered Sr$_3$Fe$_2$O$_4$Cl$_2$ with (b) the local $5$-fold pyramidal FeO$_4$Cl.}
    \label{Fig. 2}
\end{figure}

The crystal field diagram is crucial for understanding of the electronic ground state. As seen in Figures~\ref{Fig. 1} and ~\ref{Fig. 2}, tetragonal Eu$_2$SrFe$_2$O$_6$ and Sr$_3$Fe$_2$O$_4$Cl$_2$ both have bilayered structures with I4/mmm symmetry. The former has a local $4$-fold planar FeO$_2$ square lattice, and the Fe$^{2+}$ multiple $d$ states rank in the sequence ${3z^2-r^2}$ \textless~{${xz,yz}$} \textless~${xy}$ \textless~${x^2-y^2}$, as shown in Figure~\ref{Fig. 1}(c). This level sequence will be confirmed by the following density functional calculations. However, the latter forms a local $5$-fold pyramidal FeO$_2$Cl elongated along the $z$($c$) axis. Despite the different local structure and crystal field in Sr$_3$Fe$_2$O$_4$Cl$_2$, the elongated Fe-Cl bond along the $z$($c$) axis provides a similar level sequence as in Eu$_2$SrFe$_2$O$_6$.

In this work, we study the magnetic and electronic properties of Eu$_2$SrFe$_2$O$_6$ and Sr$_3$Fe$_2$O$_4$Cl$_2$ using density functional calculations and Monte Carlo simulations. We find that both materials are Mott insulators, and they have a unified HS $S$ = 2 Fe$^{2+}$ state with a $(3z^2-r^2)^2(xz,yz)^2(xy)^1(x^2-y^2)^1$ configuration. The orbital mixing between $(3z^2-r^2)$ and $(xz,yz)$ via SOC yields a small in-plane orbital moment and anisotropy. Moreover, the multiorbital superexchanges produce the strong AF couplings. And our Monte Carlo simulations give the $T_{\rm N}$ = 420 K (372 K) for Eu$_2$SrFe$_2$O$_6$ (Sr$_3$Fe$_2$O$_4$Cl$_2$), which well reproduce the experimental results. Therefore, our work provides a unified picture to illustrate the electronic and magnetic properties of Eu$_2$SrFe$_2$O$_6$ and Sr$_3$Fe$_2$O$_4$Cl$_2$.

\section{Computational Details}
Density functional theory (DFT) calculations were performed by the full-potential augmented plane waves plus local orbital code (Wien2k)\cite{blaha2001}. The optimized lattice parameters of Eu$_2$SrFe$_2$O$_6$ (Sr$_3$Fe$_2$O$_4$Cl$_2$) are $a$ = $b$ = 3.89 \AA~and $c$ = 18.80 \AA~($a$ = $b$ = 3.96 \AA~and $c$ = 22.48 \AA), which are close to the experimental values of $a$ = $b$ = 3.96 \AA~ and $c$ = 19.01 \AA~($a$ = $b$ = 4.01 \AA~ and $c$ = 22.65 \AA)~\cite{dixon2010,lopez2020}. A $\sqrt{2}$$a$ $\times$ $\sqrt{2}$$b$ $\times$ $c$ supercell is used to treat the intralayer AF configuration. The muffin-tin sphere radii are chosen to be 2.5, 2.0, 1.8, 1.5 Bohr for Eu/Sr, Fe, Cl, and O atoms, respectively. The cutoff energy of 12 Ry is used for plane-wave expansion of the interstitial wave function, and 200 $k$ points are sampled for integration over the first Brillouin zone. The generalized gradient approximation (GGA)~\cite{perdew1996} by Perdew, Burk, and Erbzerhof (PBE) is employed for the exchange-correlation potential. The SOC is included for the Eu 4$f$ and Fe 3$d$ orbitals by the second-variational method with scalar relativistic wave functions. Considering the correlation effect of Eu 4$f$ and Fe 3$d$ electrons, the GGA+SOC+$U$~\cite{anisimov1993,anisimov1997} method is employed with a common value of $U$ = 10 eV (Hund exchange $J_{\rm H}$ = 0.9 eV) for the Eu 4$f$ electrons~\cite{kovacik2016} and $U$ = 5 eV ($J_{\rm H}$ = 0.9 eV) for the Fe 3$d$ electrons~\cite{wang2014}. We examine the charge-spin state and orbital states by crystal field analyses and spin-polarized GGA calculations and then determine the electronic ground state and magnetic anisotropy by GGA+SOC+$U$, as detailed below. Moreover, we perform Monte Carlo simulations on 6$\times$6$\times$1 and 20$\times$20$\times$1 spin matrix by the Metropolis method~\cite{metropolis1949} to estimate the $T_{\rm N}$ of Eu$_2$SrFe$_2$O$_6$ and Sr$_3$Fe$_2$O$_4$Cl$_2$ using the obtained exchange parameter and magnetic anisotropy from the GGA+SOC+$U$ calculations.

\section{Results and discussion}

\subsection{Eu$_2$SrFe$_2$O$_6$}
\begin{figure}
    \includegraphics[width=8.5cm]{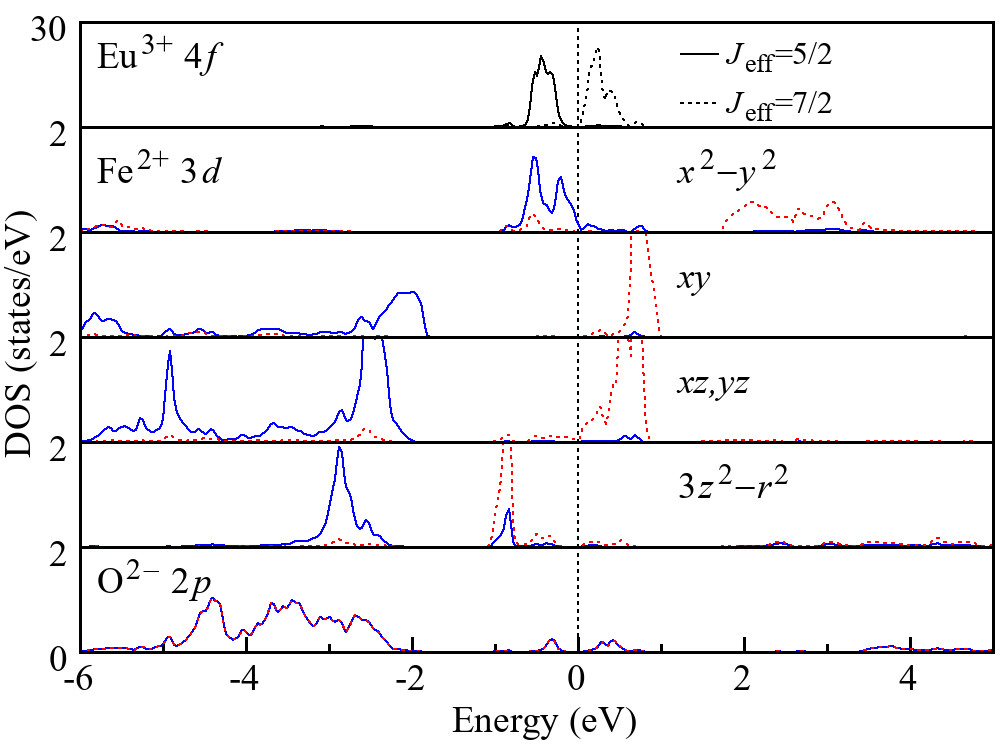}
    \caption{Eu$^{3+}$ 4$f$, Fe$^{2+}$ 3$d$, and O$^{2-}$ 2$p$ DOS results of Eu$_2$SrFe$_2$O$_6$ in the G-AF ground state by spin-polarized GGA (with inclusion of SOC only for Eu $4f$). Blue (red) curves stand for the up (down) spin channel. Fermi energy is set at zero energy.}
    \label{Fig. 3}
\end{figure}
We first examine the electronic properties of Eu$_2$SrFe$_2$O$_6$ by crystal field analyses. As seen in Figure.~\ref{Fig. 1}, four O$^{2-}$ ions surround the Fe$^{2+}$ ion to form the local $4$-fold FeO$_4$ coordination, and the corner-sharing FeO$_4$ comprise a planar square lattice. On the basis of the experimental G-AF state which is also verified by our calculations, the spin-polarized GGA calculations show that Fe 3$d$ crystal field level sequence accords with the sketched level diagram, see Figures~\ref{Fig. 1}(c) and \ref{Fig. 3}. The ${x^2-y^2}$ orbital has the highest energy and the ${xy}$ orbital follows. The ${3z^2-r^2}$ is the lowest due to the planar square FeO$_4$ structure (no apical atom). The up-spin 3$d$ orbitals are fully occupied, and the down-spin ${3z^2-r^2}$ is also occupied, suggesting the HS Fe$^{2+}$ $3d^6$ state. The O 2$p$ states lie mainly at 2-5 eV below the Fermi energy, and they have some hybridization with Fe $3d$ states. The Eu 4$f$ states, upon inclusion of the SOC, split into the lower $J$$\rm_{eff}$ = 5/2 $6$-fold states and higher $J$$\rm_{eff}$ = 7/2 $8$-fold states. The fully occupied $J$$\rm_{eff}$ = 5/2 states indicate the nonmagnetic Eu$^{3+}$ $4f^6$ state. Owing to the Fe $3d$ exchange splitting and the Eu $4f$ SOC splitting, a small energy gap is already present in the GGA framework. Note that Eu$_2$SrFe$_2$O$_6$ is a narrow band system, and its bandwidth around the Fermi level is less than 1 eV. Naturally, its insulating gap would readily be increased when the electronic correlation effect is included as shown below.
\renewcommand\arraystretch{1.5}
\begin{table}[b]
  \caption{Relative Total Energies $\Delta E$ (meV/Fe), Local Spin, and Orbital Moments($\mu_{\rm B}$) of the G-AF, C-AF and A-AF States for Eu$_2$SrFe$_2$O$_6$ by GGA+SOC+$U$.$^{a}$}
  \label{Table 1}
\setlength{\tabcolsep}{1mm}{
\begin{tabular*}{0.48\textwidth}{@{\extracolsep{\fill}}llrrrl}
\hline\hline
 & & $\Delta E$ & Fe$_{\rm spin}^{2+}$ & Fe$_{\rm orb}^{2+}$ \\ \hline
{G-AF} & $d^{5\uparrow}$$({3z^2-r^2})^\downarrow$,$\parallel$ & 0 & $\pm$3.46 & $\pm$0.07 \\
   ~  & $d^{5\uparrow}$$({3z^2-r^2})^\downarrow$,$\perp$     & 2.8 & $\pm$3.46 & $\pm$0.01 \\
  ~   & $d^{5\uparrow}$$L$$^{\downarrow}_{z+}$,$\parallel$       &879.0 & $\pm$3.36 & $\pm$0.03 \\
  ~ &  $d^{5\uparrow}$$L$$^{\downarrow}_{z+}$,$\perp$           &852.7& $\pm$3.35 & $\pm$0.80 \\ \hline
C-AF & $d^{5\uparrow}$$({3z^2-r^2})^\downarrow$,$\parallel$ & 19.3 & $\pm$3.43 & $\pm$0.08 \\
A-AF   & $d^{5\uparrow}$$({3z^2-r^2})^\downarrow$,$\parallel$ & 159.9 & $\pm$3.54 & $\pm$0.09 \\ \hline
\multicolumn{1}{l}{\textit{J}$_{ab}$=9.99} & \multicolumn{2}{c}{\textit{J}$_{c}$=4.82} &    \multicolumn{3}{c}{$D$=--0.70} ~~~\\ \hline \hline
\end{tabular*}}
$^{a}$$\parallel$ ($\perp$) represents the in-plane (out-of-plane) magnetization. The derived exchange and anisotropy parameters (meV) are given.
\end{table}

\begin{figure}[t]
    \centering
    \includegraphics[width=8.5cm]{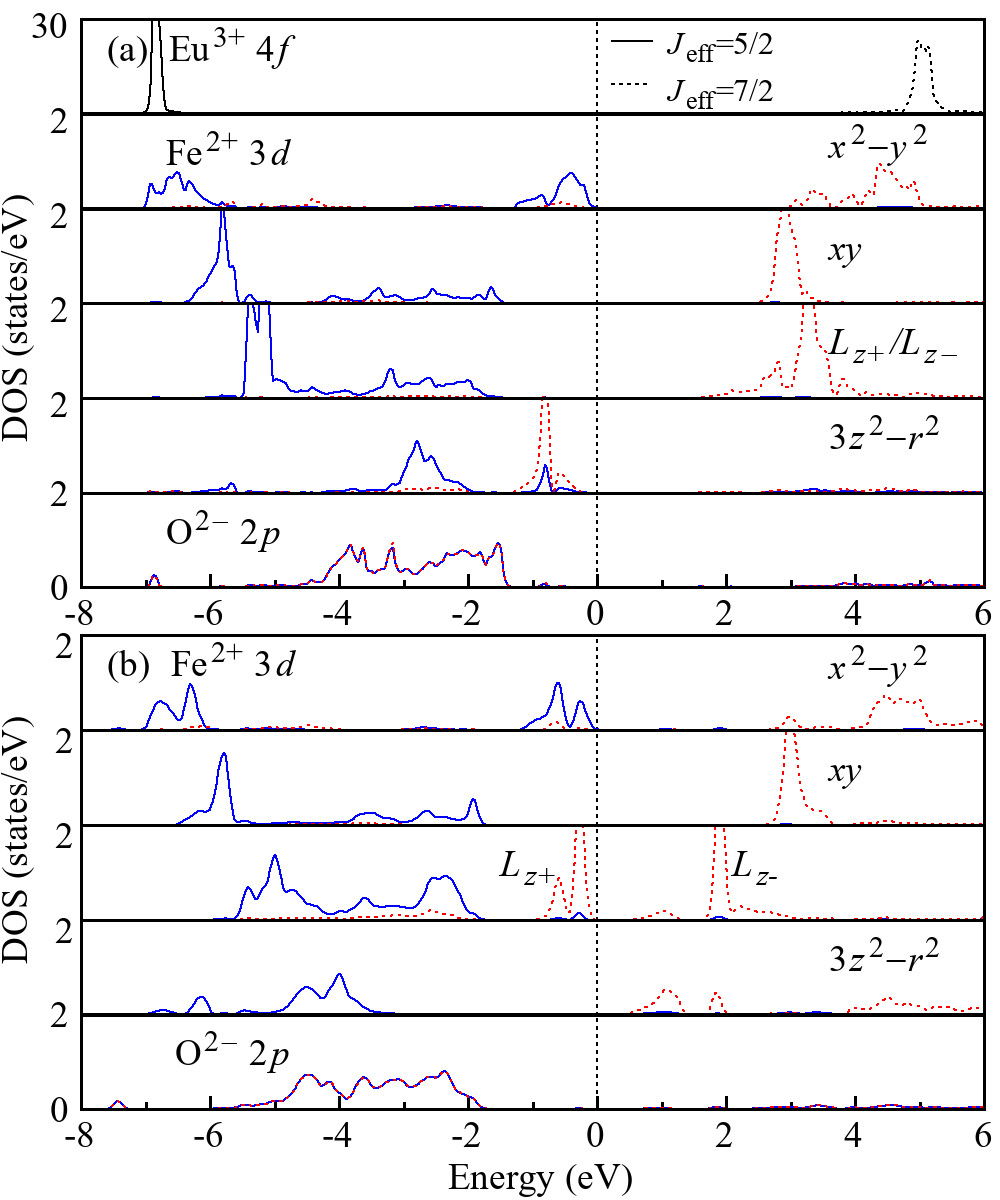}
    \caption{(a) Eu$^{3+}$ 4$f$, Fe$^{2+}$ 3$d$, and O$^{2-}$ 2$p$ DOS of the $d^{5\uparrow}$$({3z^2-r^2})^\downarrow$ ground state for the G-AF Eu$_2$SrFe$_2$O$_6$ by GGA+SOC+$U$; (b) Fe$^{2+}$ 3$d$ and O$^{2-}$ 2$p$ DOS of the $d^{5\uparrow}$$L$$^{\downarrow}_{z+}$ state. Blue (red) curves stand for the up (down) spin channel. Fermi energy is set at zero energy.}
    \label{Fig. 4}
\end{figure}

We carry out GGA+SOC+$U$ calculations to address the electronic correlation and SOC effect. To make a direct comparison among the different Fe $3d$ orbital multiplets, we initialize the corresponding density matrix (and the orbitally dependent potential) and then do a full electronic relaxation in the self-consistent way. On the basis of the G-AF state, the total-energy results of different orbital states with different spin orientations are summarized in Table~\ref{Table 1}. The Fe$^{2+}$ $d^{5\uparrow}$$({3z^2-r^2})^\downarrow$ state turns out to be the ground state, fully in line with the HS $S$ = 2 state and the crystal field level sequence with the lowest $3z^2-r^2$ orbital. It has the local spin moment of 3.46 $\mu_{\rm B}$ reduced by the Fe $3d$-O $2p$ hybridization. The DOS results are shown in Figure~\ref{Fig. 4}(a): The occupied $J$$\rm_{eff}$ = 5/2 states of the Eu$^{3+}$ $4f$ and the unoccupied $J$$\rm_{eff}$ = 7/2 states are drastically split due to the very strong electronic correlation of the highly localized 4$f$ orbitals. The Fe$^{2+}$ $d^{5\uparrow}$$({3z^2-r^2})^\downarrow$ state is also clearly shown, and this Mott insulator has a large band gap of more than 2 eV. As the Eu$^{3+}$ ions are nonmagnetic, we need to consider only the Fe$^{2+}$ ions for the magnetism of Eu$_2$SrFe$_2$O$_6$.

When the SOC effect is active, the Fe$^{2+}$ $(xz,yz)$ doublet splits into $L_{z+}$ and $L_{z-}$ states, with respective orbital moments of +1 $\mu_{\rm B}$ and --1 $\mu_{\rm B}$ along the $z$ axis (i.e., the crystallographic $c$ axis). The SOC mixing between $(3z^2-r^2)$ and $yz$ would produce the $L_{x+\sqrt{3}}$ and $L_{x-\sqrt{3}}$ states in the maximum limit with respective orbital moments of $+\sqrt{3}$ $\mu_{\rm B}$ and $-\sqrt{3}$ $\mu_{\rm B}$ along the $x$ axis
\begin{equation}
\begin{split}
&\left|L_{z\pm}\right\rangle=\frac{1}{\sqrt{2}}(\left|xz\right\rangle \mp i\left|yz\right\rangle),  \\
&\left|L_{x\pm\sqrt{3}}\right\rangle=\frac{1}{\sqrt{2}}(\left|3z^2-r^2\right\rangle \mp i\left|yz\right\rangle).\\
\end{split}
\end{equation}
As the crystal field splitting of about 1 eV between the down-spin $(3z^2-r^2)$ and $(xz,yz)$ (see Figure~\ref{Fig. 3}) is much larger than the Fe$^{2+}$ SOC strength of about 50 meV, the Fe$^{2+}$ $d^{5\uparrow}$$({3z^2-r^2})^\downarrow$ ground state carries only a small in-plane orbital moment of 0.07 $\mu_{\rm B}$ in our GGA+SOC+$U$ calculation, see Table~\ref{Table 1}. This in-plane orbital moment favors the in-plane magnetization with a magnetic anisotropy energy of 2.8 meV/Fe. This in-plane magnetization is in good agreement with both the experiment results~\cite{lopez2020} and the prediction by the selection rule based on the SOC-induced HOMO-LUMO interaction.~\cite{whangbo2015} In contrast, we also obtain the $d^{5\uparrow}$$L$$^{\downarrow}_{z+}$ state as seen in Table~\ref{Table 1} and Figure~\ref{Fig. 4}(b). This state has a large orbital moment of 0.80 $\mu_{\rm B}$ along the $z$ axis and favors a strong perpendicular magnetic anisotropy of 26.3 meV/Fe (i.e., 879.0-852.7). However, this state lies much higher than the $d^{5\uparrow}$$({3z^2-r^2})^\downarrow$ ground state by 852.7 meV/Fe due to the large crystal field excitation energy from the $3z^2-r^2$ orbital to ($xz$,$yz$).
\begin{figure}[t]
    \centering
    \includegraphics[width=6cm]{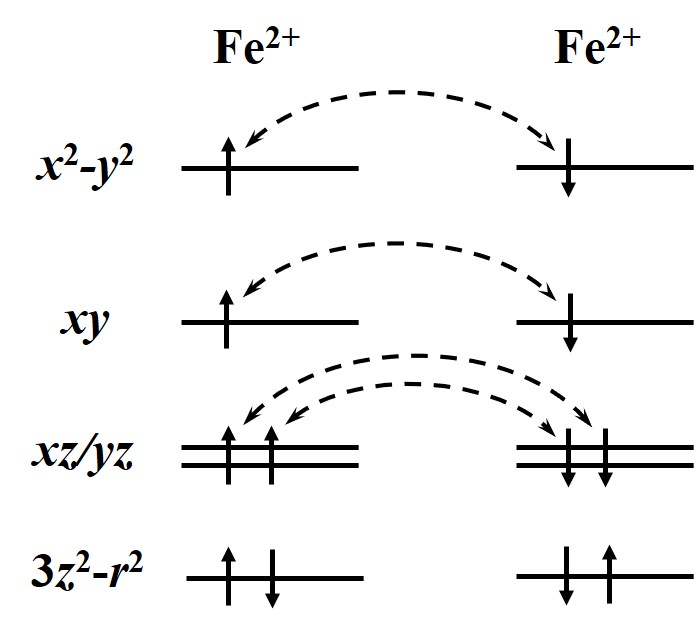}
    \caption{Schematic crystal field level diagram of Fe$^{2+}$ $S$ = 2 ions in Eu$_2$SrFe$_2$O$_6$ and Sr$_3$Fe$_2$O$_4$Cl$_2$. Virtual electron hoppings between the neighboring Fe$^{2+}$ favor strong AF superexchanges via four channels (the dashed curves).}
    \label{Fig. 5}
\end{figure}

Now we verify the experimental G-AF ground state of Eu$_2$SrFe$_2$O$_6$. On the basis of the stable $d^{5\uparrow}$$({3z^2-r^2})^\downarrow$ configuration, we calculate different magnetic configurations (A-AF with intralayer FM and interlayer AF, C-AF with intralayer AF and interlayer FM, G-AF with both intra and interlayer AF) within the GGA+SOC+$U$. As seen in Table~\ref{Table 1}, the G-AF ground state is 159.9 (19.3) meV/Fe lower than the A-AF (C-AF) state, implying the strong intralayer (relatively weak interlayer) AF coupling between the neighbouring Fe atoms. Now we explain the G-AF ground state with a high $T_{\rm N}$ through the crystal field pictures and intersite orbital interactions. For the magnetic exchange in this Mott insulator, there are four channels for the virtual electron hoppings, see Figure~\ref{Fig. 5} (the doubly occupied ${3z^2-r^2}$ orbital is magnetically inactive). The strongest $pd\sigma$ hybridization between the Fe $d$$_{x^2-y^2}$ and the O $p$$_{x,y}$ in the $ab$ plane produces the strongest intralayer superexchange AF. Another intralayer AF channel is provided by the moderate Fe $d$$_{xy}$ -- O $p$$_{x,y}$ $pd\pi$ hybridization. In addition, the Fe $d$$_{xz/yz}$ electrons contribute to the relatively weak intralayer and interlayer AF couplings. It is this multiorbital AF superexchange which accounts for the experimental high-$T_{\rm N}$ G-AF state.

\begin{figure}[t]
    \centering
    \includegraphics[width=9cm]{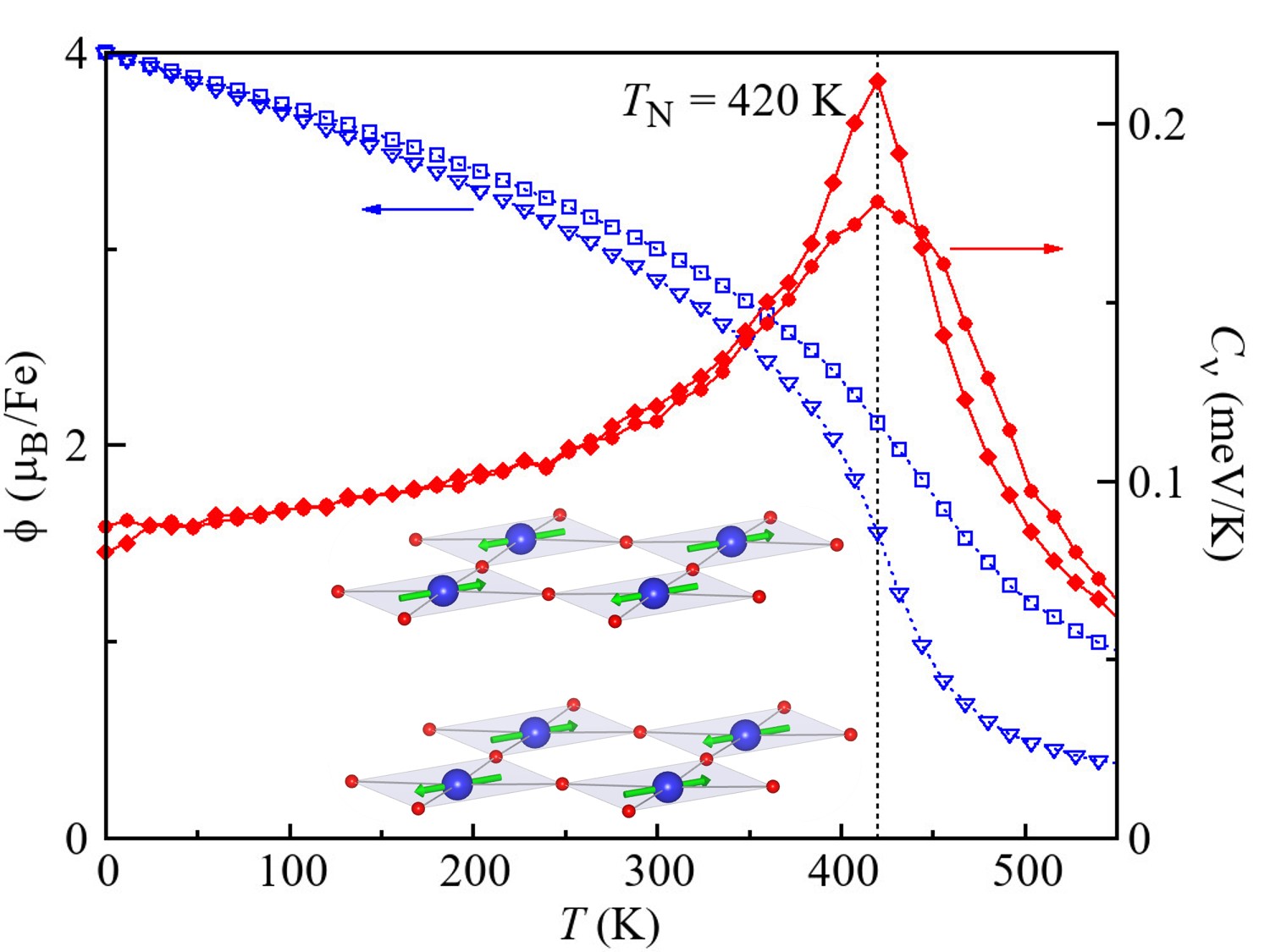}
    \caption{Monte Carlo simulations of the magnetic specific heat ($C_\nu$) of Eu$_2$SrFe$_2$O$_6$ and the AF order parameter ($\phi$) using the 6$\times$6$\times$1 spin matrix (upper $\phi$ and lower $C_\nu$ curves) and 20$\times$20$\times$1 (lower $\phi$ and upper $C_\nu$ curves). (Inset) G-AF ground state in a bilayered unit. Order parameter approaches zero at high temperature with increasing matrix sizes,~\cite{brown2005,crokidakis2009} but estimated $T_{\rm N}$ remains unchanged.}
    \label{Fig. 6}
\end{figure}
By counting the magnetic exchange energy $-JS^2$ for each AF spin pair with $S$ = 2, the exchange energies (per Fe) of the three AF states are written as follows
\begin{equation}
\begin{aligned}
E_{\rm{A-AF}} &= (2J_{ab} - 0.5J_c)S^2, \\
E_{\rm{C-AF}} &= (-2J_{ab} + 0.5J_c)S^2, \\
E_{\rm{G-AF}} &= (-2J_{ab} - 0.5J_c)S^2.
\end{aligned}
\end{equation}
Using the total energy results in Table~\ref{Table 1}, we obtain the in-plane exchange parameter $J$$_{ab}$ = 9.99 meV, and the out-of-plane exchange parameter $J$$_c$ = 4.82 meV. To probe the $T_{\rm N}$ of Eu$_2$SrFe$_2$O$_6$, we carry out Monte Carlo simulations and assume the spin Hamiltonian
\begin{equation}
H=\frac{J}{2} \sum_{\langle i j\rangle} \overrightarrow{S_{i}}\cdot \overrightarrow{S_{j}}-D \sum_{i}(S^z_i)^{2},
\end{equation}
where the first term denotes the isotropic Heisenberg exchange (AF when $J$ \textgreater~0) and the second term stands for the single-ion anisotropy [positive (negative) $D$ for the perpendicular (parallel) magnetic anisotropy]. From the above GGA+SOC+$U$ result of the easy in-plane magnetization with the anisotropy energy of 2.8 meV/Fe, $D$ = $-$0.70 meV can be derived. Using the $J_{ab}$, $J_c$, and $D$ parameters, our Monte Carlo simulations of the magnetic specific heat give $T_{\rm N}$ = 420 K for Eu$_2$SrFe$_2$O$_6$, see Figure~\ref{Fig. 6}, and it is close to the experimental magnetic specific heat of about 400 K~\cite{lopez2020}.

\begin{figure}[t]
    \includegraphics[width=8.5cm]{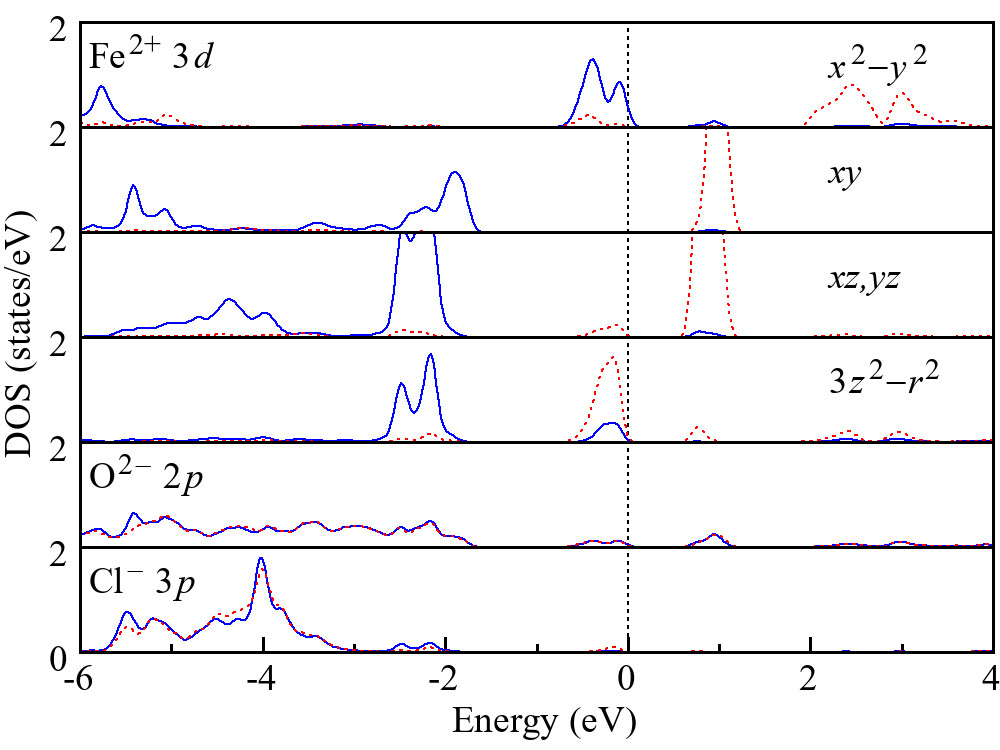}
    \caption{Fe$^{2+}$ 3$d$, O$^{2-}$ 2$p$, and Cl$^{-}$ 3$p$ DOS for the G-AF Sr$_3$Fe$_2$O$_4$Cl$_2$ by spin-polarized GGA. Blue (red) curves stand for the up (down) spin channel. Fermi energy is set at zero energy.}
    \label{Fig. 7}
\end{figure}

Moreover, to imitate the magnetization (the spontaneous magnetic response in FM materials) for the AF Eu$_2$SrFe$_2$O$_6$, we calculate the AF order parameter\cite{brown2005} (the staggered magnetization)
\begin{equation}
\phi=|\frac{1}{L^{3}} \sum_{x=1}^{L}\sum_{y=1}^{L}\sum_{z=1}^{L} (-1)^{x+y+z} \hat{s}(x,y,z)|,
\end{equation}
with $x$, $y$, and $z$ being the integer lattice coordinates of the spins. As shown in Figure~\ref{Fig. 6}, $\phi$~simulates the magnetization of the Fe$^{2+}$ ions in G-AF Eu$_2$SrFe$_2$O$_6$ with a saturated magnetic moment of 4 $\mu_{\rm B}$/Fe, and this is in line with the Fe$^{2+}$ $S$ = 2 state.

\subsection{Sr$_3$Fe$_2$O$_4$Cl$_2$}

\renewcommand\arraystretch{1.5}
\begin{table}[b]
  \caption{Relative Total Energies $\Delta E$ (meV/Fe), Local Spin, and Orbital Moments ($\mu_{\rm B}$) of the G-AF, C-AF and A-AF states for Sr$_3$Fe$_2$O$_4$Cl$_2$ by GGA+SOC+$U$.$^{a}$}
  \label{Table 2}
\setlength{\tabcolsep}{1mm}{
\begin{tabular*}{0.48\textwidth}{@{\extracolsep{\fill}}llrrrl}
\hline\hline
 & & $\Delta E$ & Fe$_{\rm spin}^{2+}$ & Fe$_{\rm orb}^{2+}$ \\ \hline
{G-AF} & $d^{5\uparrow}$$({3z^2-r^2})^\downarrow$,$\parallel$ & 0 & $\pm$3.50 & $\pm$0.09\\
 ~  & $d^{5\uparrow}$$({3z^2-r^2})^\downarrow$,$\perp$     & 2.6 & $\pm$3.50 & $\pm$0.01 \\
 ~   & $d^{5\uparrow}$$L$$^{\downarrow}_{z+}$,$\parallel$       &645.2 & $\pm$3.39 & $\pm$0.04\\
 ~ & $d^{5\uparrow}$$L$$^{\downarrow}_{z+}$,$\perp$           &616.1 & $\pm$3.39 & $\pm$0.82 \\ \hline
C-AF & $d^{5\uparrow}$$({3z^2-r^2})^\downarrow$,$\parallel$ & 27.5 & $\pm$3.45 & $\pm$0.10 \\
A-AF   & $d^{5\uparrow}$$({3z^2-r^2})^\downarrow$,$\parallel$ & 131.7 & $\pm$3.56 & $\pm$0.09 \\ \hline
\multicolumn{1}{l}{\textit{J}$_{ab}$=8.23} & \multicolumn{2}{c}{\textit{J}$_{c}$=6.88} &    \multicolumn{3}{c}{$D$=$-$0.65} ~~~\\ \hline \hline
\end{tabular*}}
$^{a}$$\parallel$ ($\perp$) represents the in-plane (out-of-plane) magnetization. The derived exchange and anisotropy parameters (meV) are given.
\end{table}

Sr$_3$Fe$_2$O$_4$Cl$_2$ has the same tetragonal I4/mmm symmetry as Eu$_2$SrFe$_2$O$_6$, but the additional $z$ axis Cl atom for the FeO$_4$ square forms a local $5$-fold pyramid. The optimized Fe-O (Cl) bondlengths of 2.02 (3.01) \AA~ are close to the experimental lengths of 2.01 (2.98) \AA~\cite{dixon2010}. The elongated FeO$_4$Cl pyramid gives a level sequence similar to the case in Eu$_2$SrFe$_2$O$_6$, see Figures~\ref{Fig. 1} and ~\ref{Fig. 2}. Indeed, we verify this picture by the spin-polarized GGA calculations, see Figures~\ref{Fig. 3} and ~\ref{Fig. 7} for a comparison.

\begin{figure}[t]
    \centering
    \includegraphics[width=8.5cm]{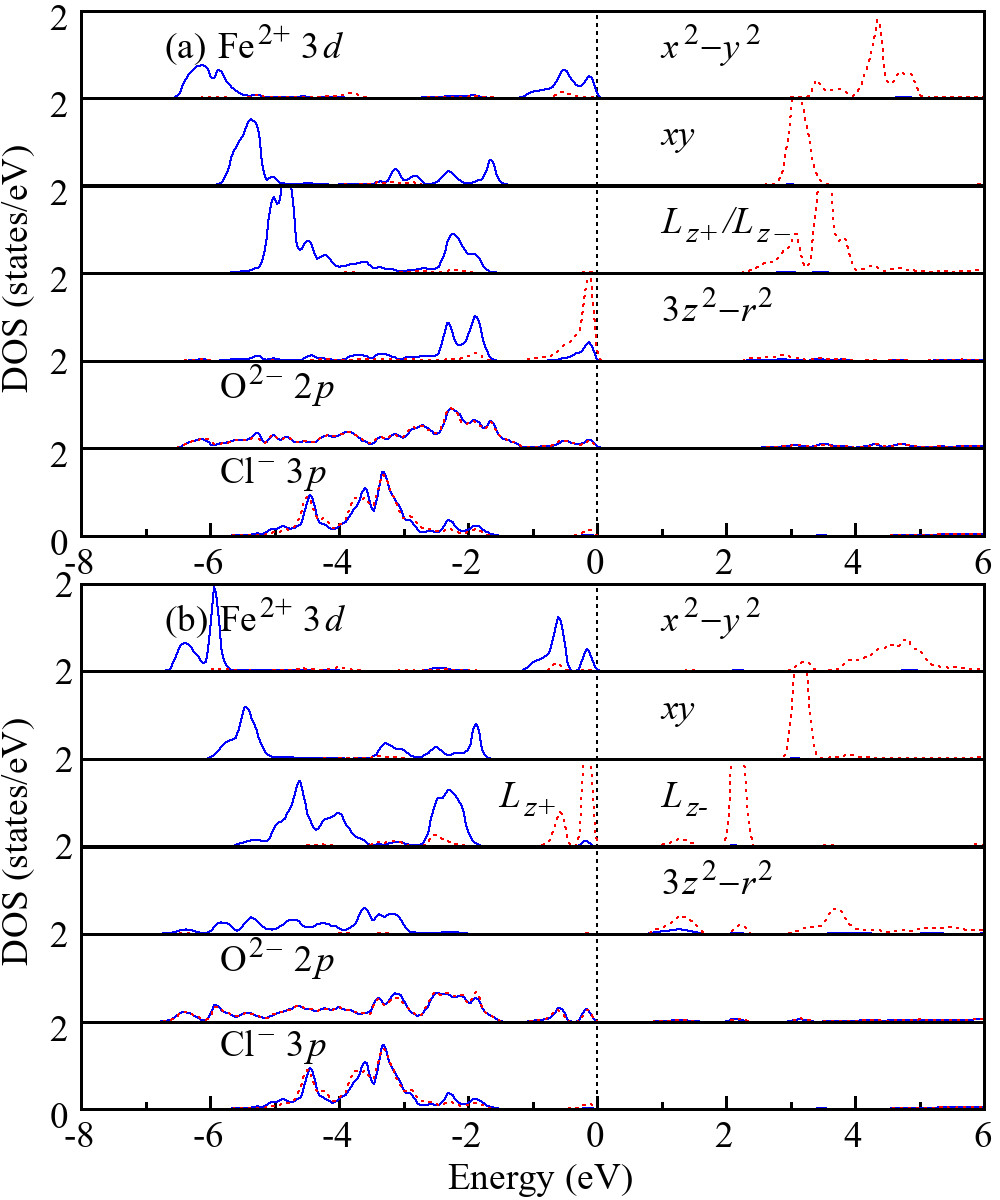}
    \caption{(a) Fe$^{2+}$ 3$d$, O$^{2-}$ 2$p$, and Cl$^{-}$ 3$p$ DOS of $d^{5\uparrow}$$({3z^2-r^2})^\downarrow$ ground state for the G-AF Sr$_3$Fe$_2$O$_4$Cl$_2$ by GGA+SOC+$U$; (b) Fe$^{2+}$ 3$d$, O$^{2-}$ 2$p$, and Cl$^{-}$ 3$p$ DOS of the $d^{5\uparrow}$$L$$^{\downarrow}_{z+}$ state . Blue (red) curves stand for the up (down) spin channel. Fermi energy is set at zero energy.}
    \label{Fig. 8}
\end{figure}

\begin{figure}[t]
    \centering
    \includegraphics[width=9cm]{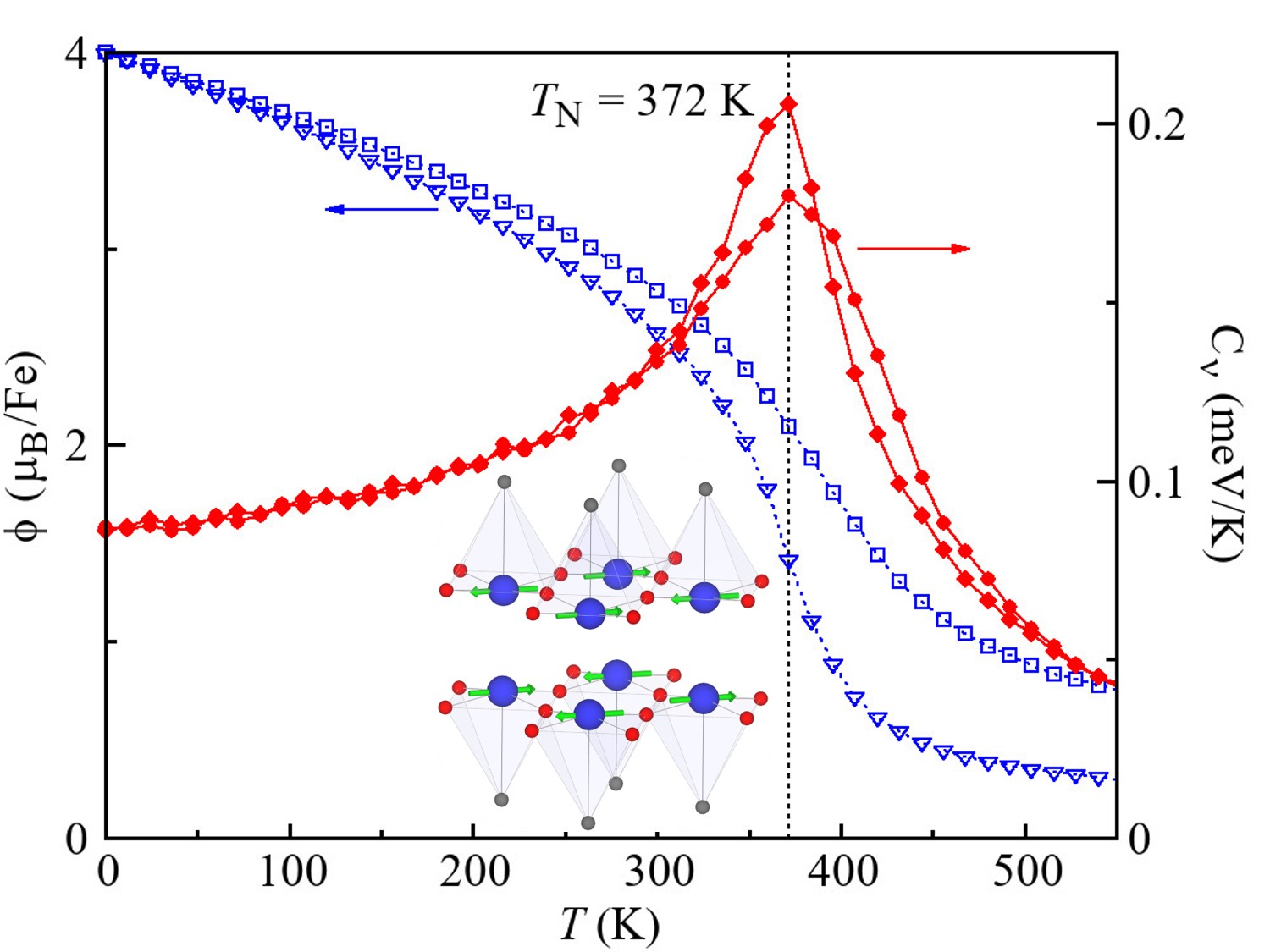}
    \caption{Monte Carlo simulations of the magnetic specific heat ($C_\nu$) of Sr$_3$Fe$_2$O$_4$Cl$_2$ and the AF order parameter ($\phi$). (Inset) G-AF ground state in a bilayered unit. See more notes in Figure~\ref{Fig. 6}.}
    \label{Fig. 9}
\end{figure}

Including the electron correlation and SOC effect, our GGA+SOC+$U$ calculations find the $d^{5\uparrow}$$({3z^2-r^2})^\downarrow$ G-AF insulating ground state with an easy in-plane magnetization in Sr$_3$Fe$_2$O$_4$Cl$_2$, see Table~\ref{Table 2} and Figure~\ref{Fig. 8}(a). It has a local spin moment of 3.50 $\mu_{\rm B}$ and a small in-plane orbital moment of 0.09 $\mu_{\rm B}$, which brings about the magnetic anisotropy of about 2.6 meV/Fe. The $d^{5\uparrow}$$L$$^{\downarrow}_{z+}$ state [see Figure~\ref{Fig. 8}(b)] is now 616.1 meV/Fe higher than the $d^{5\uparrow}$$({3z^2-r^2})^\downarrow$ ground state. This energy difference is smaller than that of 852.7 meV/Fe in Eu$_2$SrFe$_2$O$_6$, as a consequence of the lifted ${3z^2-r^2}$ level by the additional Fe-Cl bond along the $z$ axis in the former and of the reduced energy excitation from ${3z^2-r^2}$ to ($xz$,$yz$). Moreover, by comparing the total energies of the G-AF, C-AF, and A-AF states, we derive the exchange and anisotropy parameters $J$$_{ab}$ = 8.23 meV, $J$$_c$ = 6.88 meV, and $D$ = $-$0.65 meV. Then, our Monte Carlo simulations give $T_{\rm N}$ = 372 K for Sr$_3$Fe$_2$O$_4$Cl$_2$ (see Figure~\ref{Fig. 9}), and this is in good agreement with the experimental value of 378 K~\cite{dixon2010}.

Previously, it was found that the bilayered Sr$_3$Fe$_2$O$_4$Cl$_2$ has a perpendicular magnetic moment of 2.82 $\mu_{\rm B}$/Fe~\cite{dixon2010}. We now check this observation carefully. Indeed, we obtain the $d^{5\uparrow}$$L$$^{\downarrow}_{z+}$ solution with perpendicular magnetization, and it has the spin moment of 3.39 $\mu_{\rm B}$ per Fe and perpendicular orbital moment of 0.82 $\mu_{\rm B}$. The total magnetic moment of 4.21 $\mu_{\rm B}$ is much larger than the above value of 2.82 $\mu_{\rm B}$. However, this solution is significantly higher (by 616.1 meV/Fe) than the $d^{5\uparrow}$$({3z^2-r^2})^\downarrow$ ground state with a small in-plane orbital moment as discussed above. Therefore, the previous observation of the perpendicular magnetic moment of 2.82 $\mu_{\rm B}$/Fe in Sr$_3$Fe$_2$O$_4$Cl$_2$ cannot be explained by the present finding. Note that from symmetry analysis,~\cite{whangbo2015} the experimental perpendicular magnetization would most probably arise from the Fe$^{2+}$ $d^{5\uparrow}$($xz$,$yz$)$^{1\downarrow}$ state with the formal $S_z$=2 and $L_z$=1, but the total moment of 5 $\mu_{\rm B}$ is significantly larger than the experimental value of 2.82 $\mu_{\rm B}$. Therefore, the experimental perpendicular magnetic moment~\cite{dixon2010} of 2.82 $\mu_{\rm B}$ itself seems to contain inconsistent information, and it was already deemed a contrast to the in-plane magnetization of Eu$_2$SrFe$_2$O$_6$ in another very recent experimental work~\cite{lopez2020} which leaves this issue open. A similar issue appears recently for several cobaltates containing the HS Co$^{2+}$ ions at the axially elongated octahedral sites:~\cite{koo2020} a large orbital moment as reported from experiment but a small in-plane orbital moment as predicted by theory. All of these issues may be settled by a further study. It is important to note that our present work provides a unified picture for the electronic and magnetic properties of the bilayered Sr$_3$Fe$_2$O$_4$Cl$_2$ and Eu$_2$SrFe$_2$O$_6$ concerning the crystal field level diagram, superexchange AF, easy in-plane magnetization, and high-$T_{\rm N}$ G-AF order.

\section{Conclusions}

In summary, using a set of DFT calculations aided with crystal field and superexchange analyses we find that both Eu$_2$SrFe$_2$O$_6$ and Sr$_3$Fe$_2$O$_4$Cl$_2$ are Mott insulators and in the HS $S$ = 2 Fe$^{2+}$ state. Although they have different local structures and crystal fields, being $4$-fold square in the former and $5$-fold pyramid in the latter, we find the same level sequence and the Fe$^{2+}$ $d^{5\uparrow}$$({3z^2-r^2})^\downarrow$ ground state with the easy in-plane magnetization. The multiorbital superexchanges well explain the strong AF couplings in the G-AF ground state. Our Monte Carlo simulations give a high $T_{\rm N}$ = 420 K (372 K) for Eu$_2$SrFe$_2$O$_6$ (Sr$_3$Fe$_2$O$_4$Cl$_2$), which well matches the experimental value of $\sim$ 400 K (378 K). Therefore, this work provides a unified picture for the electronic structure and magnetism of Eu$_2$SrFe$_2$O$_6$ and Sr$_3$Fe$_2$O$_4$Cl$_2$.

\section{acknowledgement}

This work was supported by National Natural Science Foundation of China (Grants No. 12174062 and No. 12104307).

\bibliography{Layered_Fe_oxides}

\end{document}